\documentclass[conference]{IEEEtran}
\IEEEoverridecommandlockouts
\usepackage{cite}
\usepackage{amsmath,amssymb,amsfonts}
\usepackage{bm}
\usepackage{algorithmic}
\usepackage{graphicx}
\usepackage{textcomp}
\usepackage{xcolor}
\usepackage{float}
\usepackage{amsthm}
\newtheorem{theorem}{Theorem}  %跟随二级标题排序

\usepackage{hyperref}
\usepackage{booktabs}
\usepackage{graphicx}
\usepackage{bbding}
\usepackage{threeparttable}

\usepackage{multirow}
\usepackage{graphicx}
\usepackage{amsfonts}
\usepackage{mathtools}
\usepackage{amssymb}
\usepackage{amsmath}
\usepackage{amsthm}
\usepackage{color}
\usepackage[ruled,linesnumbered]{algorithm2e}
\usepackage{booktabs}
\usepackage{enumerate}
\usepackage{siunitx}
\usepackage{hyperref}
\usepackage{MnSymbol}
\usepackage{subfigure}
\usepackage{bm}

\makeatletter
\newcommand{\linebreakand}{%
  \end{@IEEEauthorhalign}
  \hfill\mbox{}\par
  \mbox{}\hfill\begin{@IEEEauthorhalign}
}
\makeatother

\def\BibTeX{{\rm B\kern-.05em{\sc i\kern-.025em b}\kern-.08em
    T\kern-.1667em\lower.7ex\hbox{E}\kern-.125emX}}
\begin{document}

\title{PROFL: A Privacy-Preserving Federated Learning Method with Stringent Defense Against Poisoning Attacks
}

\author{
\IEEEauthorblockN{1\textsuperscript{st} Yisheng Zhong, 2\textsuperscript{nd} Li-Ping Wang$^*$}

\IEEEauthorblockA{\textit{State Key Laboratory of Information Security, Institute of Information Engineering, Chinese Academy of Sciences} \\
\textit{School of Cyber Security, University of Chinese Academy of Sciences, 
Beijing, China} \\
\{zhongyisheng,wangliping\}@iie.ac.cn}
\linebreakand

% \IEEEauthorblockN{2\textsuperscript{nd} Li-Ping Wang}
% \IEEEauthorblockA{\textit{State Key Laboratory of Information Security, Institute of Information Engineering, CAS} \\
% \textit{School of Cyber Security, University of Chinese Academy of Sciences}\\
% Beijing, China \\
% wangliping@iie.ac.cn}
}

\maketitle
\begin{abstract}
Federated Learning~(FL)~faces two major issues:~privacy leakage and poisoning attacks, which may seriously undermine the reliability and security of the system. Overcoming them simultaneously poses a great challenge. This is because privacy protection policies prohibit access to users' local gradients to avoid privacy leakage, while Byzantine-robust methods necessitate access to these gradients to defend against poisoning attacks. To address these problems, we propose a novel privacy-preserving Byzantine-robust FL framework PROFL. PROFL is based on the two-trapdoor additional homomorphic encryption algorithm and blinding techniques to ensure the data privacy of the entire FL process. During the defense process, PROFL first utilize secure Multi-Krum algorithm to remove malicious gradients at the user level. Then, according to the Pauta criterion, we innovatively propose a statistic-based privacy-preserving defense algorithm to eliminate outlier interference at the feature level and resist impersonation poisoning attacks with stronger concealment. Detailed theoretical analysis proves the security and efficiency of the proposed method. We conducted extensive experiments on two benchmark datasets, and PROFL improved accuracy by 39\% to 75\% across different attack settings compared to similar privacy-preserving robust methods, demonstrating its significant advantage in robustness.
\end{abstract}


\section{Introduction}
Federated Learning (FL) allows data owners to enhance the efficiency and accuracy of global models without exposing their private data. FL has broad potential applications in fields such as healthcare~\cite{dayan2021federated} and finance~\cite{DBLP:conf/ijcai/ZhengYG020}. However, FL also faces problems including privacy leakage and poisoning attacks, which pose a threat to its security and reliability.

Privacy leakage that results in data security issues is the first major threat faced by FL. As the gradients contain some private information, uploading them directly to the central server without undergoing special treatment can result in leakage issues~\cite{DBLP:conf/nips/ZhuLH19}. Poisoning attack leading to a decrease in model reliability is the another major threat to FL. Byzantine adversaries employ poisoning attacks to manipulate malicious participants and intentionally inject harmful information into the uploaded gradient. This manipulation leads to gradient deviation, ultimately undermining model performance~\cite{DBLP:conf/nips/BlanchardMGS17} and altering model decisions~\cite{DBLP:conf/iclr/XieHCL20}. Therefore, developing a FL mechanism that simultaneously preserves privacy and resists poisoning attacks is crucial to making FL a trustworthy and reliable technique.

However, there are still two important obstacles to achieving such a FL system. One is that privacy-preserving FL aims to guarantee the indistinguishability of the data, while defensive approaches necessitate access to local gradients and use similarity measures to distinguish malicious gradients from benign gradients, which seems paradoxical. The other is that there are now cunning Byzantine adversaries who make poisoning gradient attacks more covert. For example, some camouflage methods~\cite{DBLP:conf/nips/BaruchBG19} are capable of analyzing the range of parameter changes that the defender cannot detect and then evade defense mechanisms. Recent approaches achieve a balance between privacy and robustness. For instance, PEFL~\cite{DBLP:journals/tifs/LiuLXCHL21} utilizes the Pearson correlation coefficient to identify outliers, employing a dual-server model for secure computation where trusted servers hold keys to collaborate with additive homomorphic encryption algorithms. The ShieldFL~\cite{ShieldFL} adopts Two-trap-door homomorphic encryption technology to relax the trust requirement for servers, using cosine similarity to identify malicious gradients. However, their robustness and privacy are not ideal.

To address the above issues, we proposes a composite privacy-preserving Byzantine-robust algorithm called PROFL. It performs two-trapdoor additional homomorphic algorithm (AHE)~\cite{liu2016efficient} and blinding techniques on two non-colluding servers to prevent privacy leakage during the defense process. In addition, PROFL first uses the multi-krum algorithm~\cite{DBLP:conf/nips/BlanchardMGS17} based on the Euclidean distance as the similarity measure to detect and eliminate poisoning gradients at the user level. Then, PROFL uses statistical methods based on the Pauta criterion to eliminate outliers at the feature level, thereby resisting concealed impersonation poisoning attack. Such a composite defense can defend against malicious gradient attacks from macro and micro perspectives simultaneously. The main contributions are summarized as follows.

\begin{itemize}
    \item We propose a composite robust mechanism to counter both general and concealed poisoning attacks. Across multiple attack scenarios, compared to similar approaches, our method demonstrates an improvement of 39\% to 75\% in performance. This showcases the broader and more rigorous advantages of our composite defense.
    \item Compared to similar methods, both the security and efficiency of our privacy-preserving strategy are higher. PROFL effectively mitigates a broader range of collusion attacks and, unlike multi-key methods, significantly reduces the computational overhead associated with key pair generation and joint decryption.
    \item Theoretical analysis demonstrates that PROFL performs well in managing communication and time overheads.
\end{itemize}

The rest of this paper is organized as follows. In §\ref{Problem Statement}, we introduce the problem statement. In §\ref{Framework}, we present the PROFL in detail. Subsequently, we perform the theoretical analysis in §\ref{Theoretical Analysis} and evaluate the performance in §\ref{Performance Evaluation}. Finally, the paper is concluded in §\ref{Conclusion}.

\section{Problem Statement}\label{Problem Statement}

\subsection{System Model}

\begin{figure}[htbp]
  \centering
  \includegraphics[width=0.9\linewidth]{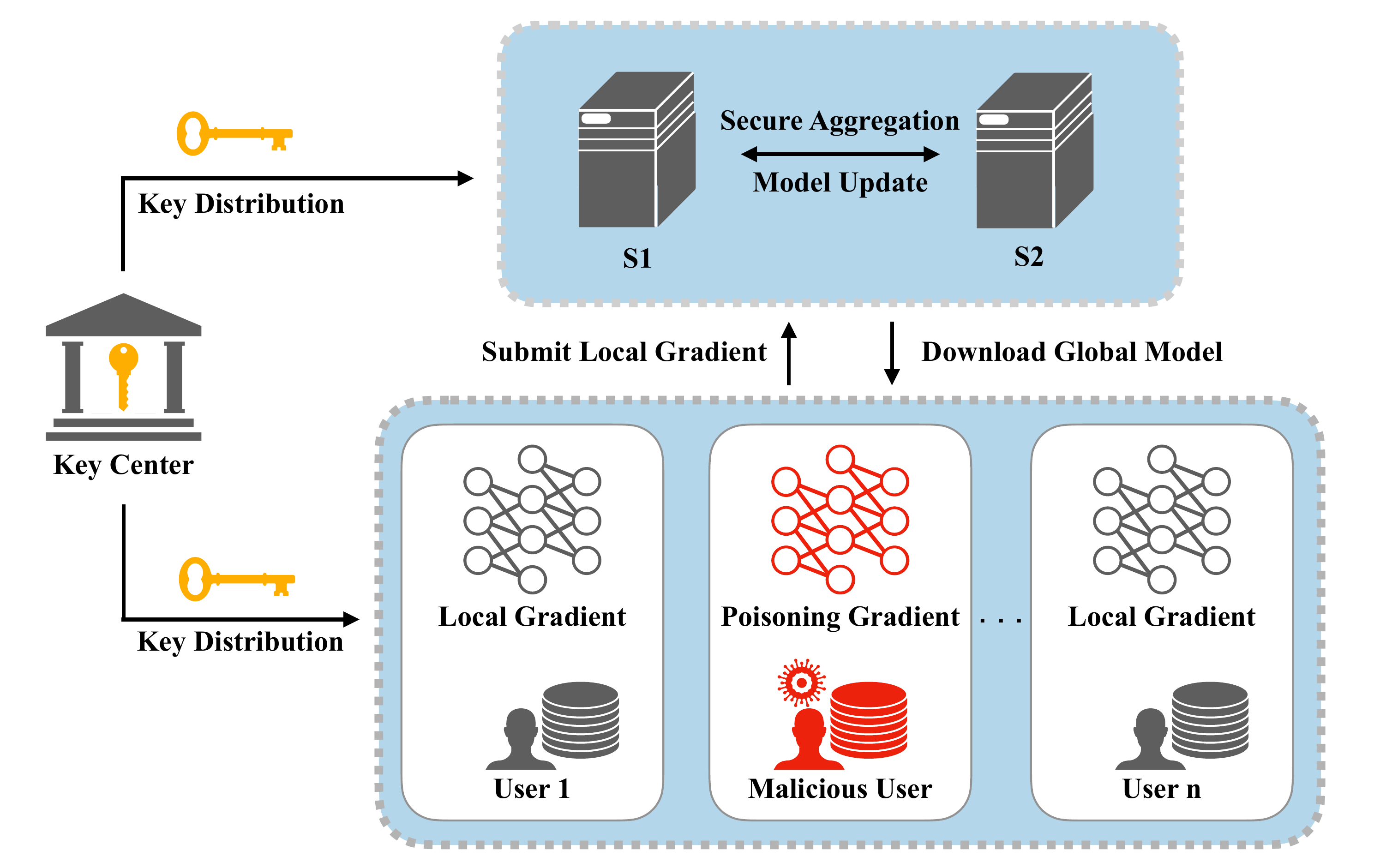}
  \caption{System Model.}
  \label{fig:system}
\end{figure}

Our system consists of a key center, a pair of servers, and multiple users. Their relationship is as shown in the Fig \ref{fig:system}.

\textbf{\textit{Key Center($KC$):}} KC is a trusted agency responsible for the generation and distribution of keys. First, it generates two public-secret key pairs and splits one of the secret keys into two shares. After generation, it distributes public and secret keys and key shares to the designated participants.

\textbf{\textit{Server($S_1$ and $S_2$):}} The servers $S_1$ and $S_2$ cooperate and restrict each other to complete complex operations over the encrypted data and resist poisoning attacks. $S_1$ is the primary server, which is responsible for communicating with users and executing robust algorithms. $S_2$ is responsible for assisting $S_1$ to complete the secure computation.

\textbf{\textit{User($U$):}} Users use the latest global model to locally train the model on private data. Then, the gradient is encrypted and uploaded to $S_1$. After $S_1$ and $S_2$ cooperate to defend and update the global model, users download the latest global model to replace their local models.

\subsection{Threat Model}

This work focuses on the threats to privacy and model robustness. We assume KC is completely trustworthy, so the possible threats come from servers and users. Firstly, the two servers are semi-honest. They both follow the pre-defined protocol, but they attempt to decrypt the users' gradients, and they cannot collude with each other~\cite{mohassel2017secureml}. Similarly, users are also semi-honest, they all follow the pre-defined protocol, but they try to collude with $S_1$ to decrypt other users' gradients. 

Secondly, the robustness of the model is threatened by malicious users. We assume that malicious users can have knowledge of benign users’ gradient distributions and other information in advance. Malicious users submit poisoning gradients to deviate the global model from the training target. Moreover, considering that defense techniques often utilize the principle of similarity to resist poisoning attacks, malicious adversaries may employ countermeasures. These adversaries cunningly design malicious gradients to be as close as possible to benign gradients in terms of distance or similarity in order to evade defenses. They also aim to make significant changes in a few dimensions of these gradients to achieve their poisoning objectives. Specifically, a malicious user will choose $p$ parameters from its malicious gradient $\mathbf{g}_m$ and modify them to arbitrarily large values of $t$ such that:
\begin{equation*}
\mathsf{Max}(\mathsf{Dis}(\mathbf{g}_i,\mathbf{g}_j) (i,j\in(1,n))) > \mathsf{Dis}(\mathbf{g}_m=(..., t_1, ..., t_p, ...), \mathbf{g}_b)
\end{equation*}
where $\mathbf{g}_b$ is a benign gradients and $\mathsf{Dis}$ is a distance function. To ensure the practicability and realizability of the defense strategy, we assume that the upper limit of malicious users is $\lfloor\frac{n}{2}\rfloor$ of all users.

\section{Framework}\label{Framework}

 In this section, we provide a detailed description of the construction of PROFL. 
 %The symbols that appeared in this paper and corresponding descriptions are listed in Table \ref{tab:symbol}.

% \begin{table}[h]
% \caption{Symbols}
% \label{tab:symbol}
% \begin{tabular}{cl}
% \toprule
% Symbols       & Descriptions                                                                \\ \midrule
% $pk, sk$        & Public/secret key                                                           \\
% $k$             & Security parameter                                                          \\
% $[\![m]\!]$ & Ciphertext                                                                  \\
% $[m]_{sk_i}$       & Decryption share with sk\_i                                                 \\
% $w$             & Weight of model                                                             \\
% $deg$           & Precision degree                                                            \\
% $\mathsf{SecDis}$        & Secure distance                                                             \\
% $\mathsf{SecSel}$        & Secure select                                                               \\
% $\mathsf{SecRep}$        & Secure represent                                                            \\
% $\mathbf{g}$             & Gradient                                                                    \\
% $\mathbf{g}_{agg}$        & Aggregated gradient                                                         \\
% $med_t$        & \begin{tabular}[c]{@{}l@{}}the t-th dimension of the $\mathbf{g}_{agg}$\end{tabular} \\ \bottomrule
% \end{tabular}
% \end{table}

In PROFL, we employ the Two-trapdoor AHE algorithm to preserve privacy, which is a type of distributed public-key cryptosystem. It is a variant of the Paillier Cryptosystem~\cite{paillier1999public} and possesses additive homomorphic properties. It mainly consists of the components $\mathsf{KeyGen}$, $\mathsf{Enc}$, $\mathsf{FullDec}$, $\mathsf{KeySplit}$, $\mathsf{PartDec1}$ and $\mathsf{PartDec2}$. Besides, the two-trapdoor AHE algorithm satisfies the additive homomorphism, and we define their operator symbols as $\oplus$ and $\odot$ respectively. Therefore, given two ciphertexts $[\![x_1]\!]$ and $[\![x_2]\!]$ and an integer $A \in {Z_N}$, we can perform such a secure calculation in the encrypted state:
\begin{align}
    [\![x_1]\!]\oplus[\![x_2]\!] = [\![x_1]\!]\cdot[\![x_2]\!]= [\![x_1+x_2]\!]\label{he1}\\ 
    [\![x_1]\!]\odot{A} = [\![x_1]\!]^{A}
    = [\![x_1 \cdot A]\!]\label{he2}
\end{align}

\subsection{Initialization}

\textbf{Key Generation and Distribution:} Firstly, KC generates two sets of public-secret keys using $(pk_m, sk_m)\gets \mathsf{KeyGen}(1^{k_1})$ and $(pk_g, sk_g)\gets \mathsf{KeyGen}(1^{k_2})$, then delivers the public key $pk_m$ to $S_1$ and broadcasts the secret key $sk_m$ to all users, which are respectively used for encrypting and decrypting the new global model. Next, KC broadcasts the public key $pk_g$ to all users for encrypting the local gradients submitted to the server. KC delivers the public key $pk_g$ to $S_1$ to encrypt the intermediate variables for secure computation. KC then splits the secret key $sk_g$ into two secret shares $sk_{g_1}$ and $sk_{g_2}$ using $(sk_{g_1}, sk_{g_2})\gets \mathsf{KeySplit}(sk_g)$, and distributes them to servers $S_1$ and $S_2$, respectively, for jointly decrypting the gradients submitted by the users.

\noindent\textbf{Model Initialization and Distribution:} Firstly, the server $S_1$ randomly initializes a global model parameter $\mathbf{w}_0$, and then uses the public key $pk_m$ to encrypt it as $[\![\mathbf{w}_0]\!]$ using $[\![\mathbf{w}_0]\!] \gets \mathsf{Enc}(pk_m, \mathbf{w}_0)$. Then $S_1$ broadcasts the encrypted model $[\![\mathbf{w}_0]\!]$ to each user. Each user $U_i$ decrypt the global model to obtain $\mathbf{w}_0$ with the secret key $sk_m$ using $\mathbf{w}_0 \gets \mathsf{FullDec}(sk_m, [\![\mathbf{w}_0]\!])$.

\subsection{Local Training}

During the gradient generation phase, Benign users utilize local data to conduct mini-batch local training on the latest model, and then accumulate the obtained gradients. In the same phase, malicious users will not submit gradients obtained by training on real data. Instead, they upload their own designed malicious gradients. As two-trapdoor AHE algorithm are operated on integers, we use the precision parameter $deg$ to convert all parameters of gradients from floating-point to integer format such that: $g' = \lfloor{g \cdot deg}\rceil$. Then, each user encrypts the gradient and submits it to $S_1$ for aggregation.

\subsection{Privacy-Preserving Defense Strategy}

To avoid the risk of a single untrusted server leaking secret keys or decrypting users' gradients, PROFL utilizes a combination of two-trapdoor AHE mechanism and blinding techniques to execute defensive strategies on two servers without revealing plaintext gradients. Furthermore, in order to prevent malicious users from colluding with either of the two servers to decrypt gradients submitted by other users, we set up two sets of public-secret key pairs. This enables users to encrypt their own submitted gradients, while preventing them from decrypting gradients submitted by other users.

To defend against poisoning attacks, PROFL proposes a secure defense algorithm based on the Multi-Krum algorithm. During the defense process, the secure distance function $\mathsf{SecDis}$ is used to calculate the Euclidean distance between two encrypted gradients. Then, the secure selection function $\mathsf{SecSel}$ selects $\lceil\frac{n}{2}\rceil$ gradients that are closest to other local gradients in geometric space as representatives. Based on the proportion of malicious users, these representative gradients are closer to the center of the group, and it is believed that they can represent the characteristics of the vast majority of benign user gradients, thereby enabling the global gradient to converge towards the optimal aggregated gradient.

To address the issue of cunning malicious users whose gradients cannot be distinguished using defense methods based on similarity principles, we propose the Secure Represent function $\mathsf{SecRep}$ to mitigate this vulnerability. $\mathsf{SecRep}$ applies the Pauta criterion to remove outliers in each dimension and then selects the median as the representative for that dimension. Through these operations, we obtain a representative global gradient. The procedure is shown in Algorithm \ref{algo:PPDS}, and the three security protocols are as follows:

\begin{algorithm}[h]
\small
	\caption{Privacy-preserving Defense Strategy}
	\label{algo:PPDS}   
	\KwIn{The encrypted local gradients $\{[\![\mathbf{g}_1]\!], [\![\mathbf{g}_2]\!], ... [\![\mathbf{g}_n]\!]\}$.}
	\KwOut{Representative gradient $[\![\mathbf{g}_{agg}]\!]$.}
	
	/*Calculate the sum of the distances of each gradient to the other gradients*/
	
	\For {$i \in [1,n]$}{$sum_i=0$;}
	% $sum_{i \in [1,n]}=0$;
 
	\For {$i \in [1,n]$}{
	    \For {$j \in [i,n]$}{
	        $dis(i,j)=\mathsf{SecDis}([\![\textbf{g}_i]\!],[\![\textbf{g}_j]\!]);$
            \mbox{/*Accumulate $dis(i,j)$ using homomorphic addition*/} \\
            $[\![sum_i]\!] = [\![sum_i]\!]\oplus dis(i,j)$, \\
            $[\![sum_j]\!] = [\![sum_j]\!]\oplus dis(i,j)$;
	    }
	}
	/*Screening Representatives*/
	
	Using $\mathsf{SecSel}$, selects $\lceil\frac{n}{2}\rceil$ gradients with the largest corresponding value in  $(sum_i)$ $(i \in [1,n])$ among n gradients, as the representative of benign gradients;

	/*Remove Outliers*/
	
	Arrange all gradients into sets $\{ d_i\}_{i \in [1,m]}$ by dimension, where $d_i=\{[\![\mathbf{g}_1[i]]\!], [\![\mathbf{g}_2[i]]\!], ...[\![\mathbf{g}_n[i]]\!]\}$ (here we use $\mathbf{g}[i]$ to denote the $i$-th element of $\mathbf{g}$);

    \For{$i \in [1:m]$}{
        Use $\mathsf{SecRep}$ algorithm to remove outliers in set $d_i$ and calculate the median $med_i$ of set $d_i$ as a representative of the $i$-th dimension;
    }
    
    \textbf{return:} aggregated gradient $[\![\textbf{g}_{agg}]\!] = [\![ ( med_1, med_2, ...med_m ) ]\!]$.
    
\end{algorithm}

\textbf{\textit{SecDis:}} $\mathsf{SecDis}$ employs additive homomorphic algorithm and blinding technique to calculate the Euclidean distance between two encrypted gradients. Given two encrypted gradients $[\![\mathbf{g}_x]\!]$ and $[\![\mathbf{g}_y]\!]$, the servers $S_1$ and $S_2$ jointly calculate the Euclidean distance $dis(i,j) \gets \mathsf{SecDis}([\![\mathbf{g}_x]\!],[\![\mathbf{g}_y]\!])$. The specified procedure is demonstrated in Fig.\ref{fig:SecDis}.

\begin{figure}[h]
  \centering
\small
    \begin{tabular}{|p{0.96\columnwidth}|}
    \hline

\begin{center} \textbf{Implementation of $\mathsf{SecDis}$}\end{center}

\textbf{Input:} $S_1$ holds two gradients $[\![\mathbf{g}_x]\!]$, $[\![\mathbf{g}_y]\!]$, the public key $pk_g$ and the secret key $sk_1$; $S_2$ holds the secret key $sk_2$. 

\textbf{Output:} Encrypted Euclidean distance $[\![d]\!]$ between $[\![\mathbf{g}_x]\!]$ and $[\![\mathbf{g}_y]\!]$.

\textbf{Procedure:}
\begin{itemize}

\item[$\bullet$] @$S_1$: To protect privacy of gradients, for $i \in \{x, y\}$, $S_1$ randomly selects two nonzero noise vectors $\mathbf{r}_i \gets Z_N^*$ to blind the gradients using Equation (\ref{he2}) such that $[\![\mathbf{g}_i']\!]=[\![\mathbf{g}_i]\!]\cdot[\![\mathbf{r}_i]\!]=[\![\mathbf{g}_i+\mathbf{r}_i]\!]$. Then, $S_1$ partially decrypts the gradient by calling 
$[\mathbf{g}_i']_{sk_1} \gets \mathsf{PartDec1}(sk_1, [\![\mathbf{g}_i']\!])$, and send {$[\mathbf{g}_i]_{sk_1}$} and {$[\![\mathbf{g}_i']\!]$} to $S_2$.

\item[$\bullet$]@$S_2$: To fully decrypt gradients, $S_2$ calls algorithm $\mathbf{g}_i' \gets \mathsf{PartDec2}(sk_2, [\![\mathbf{g}_i']\!], [\mathbf{g}_i]_{sk_1})$ $(i \in \{x,y\})$. Then $S_2$ calculates the Euclidean distance $\mathbf{dis}'$ between $\mathbf{g}_x'$ and $\mathbf{g}_y'$ such that $\mathbf{dis}'=(\mathbf{g}_x'-\mathbf{g}_y')^2=[(\mathbf{g}_x+\mathbf{r}_x)-(\mathbf{g}_y+\mathbf{r}_y)]^2$. Then $S_2$ sends $\mathbf{dis}'$ to $S_1$.

\item[$\bullet$]@$S_1$: To remove the random noises and get $[\![\mathbf{dis}]\!]$ using Equation (\ref{he1}) such that:
\begin{equation}
[\![\mathbf{dis}]\!]=[\![\mathbf{dis}'+\mathbf{t}]\!]=[\![\mathbf{dis}']\!]\cdot[\![\mathbf{t}]\!]\label{D}
\end{equation}
which needs intermediate results $[\![\mathbf{dis}']\!]$ and $[\![\mathbf{t}]\!]$. $[\![\mathbf{t}]\!]$ can be obtained using Equation (\ref{he1}) and (\ref{he2}) such that:
\begin{eqnarray}
\label{T}
& & [\![\mathbf{t}]\!]=[\![\mathbf{dis}-\mathbf{dis}']\!]=[\![(\mathbf{g}_x-\mathbf{g}_y)^2-(\mathbf{g}_x'-\mathbf{g}_y')^2]\!]  \nonumber \\
& &=[\![\mathbf{g}_{x}]\!]^{-2r_{x}} \cdot [\![-{r_x}^2]\!] \cdot [\![\mathbf{g}_{x}]\!]^{2r_{y}} \cdot [\![\mathbf{g}_{y}]\!]^{2r_{x}} \cdot [\![2r_{x}\mathbf{r}_{y}]\!]\nonumber\\
& &\quad \cdot[\![\mathbf{g}_{y}]\!]^{-2r_{y}} \cdot [\![-{r_y}^2]\!]
\end{eqnarray}

where $S_1$ owns $[\![\mathbf{g}_{x}]\!]$, $[\![\mathbf{g}_{y}]\!]$, $\mathbf{r}_{x}$ and $\mathbf{r}_{y}$. Then, removes the random noises $\mathbf{r}_x$ and $\mathbf{r}_y$ and gets $[\![\mathbf{dis}]\!]$ using Equation (\ref{D}), (\ref{T}) and $[\![\mathbf{dis}']\!] \gets \mathsf{Enc}(pk_g, \mathbf{dis}')$. Finally, accumulate each dimension in $\mathbf{dis}$ to obtain a numerical value $[\![d]\!]$ for the distance between gradients using Equation (\ref{he1}) such that $\prod_{i=1}^{m}[\![\mathbf{dis}(i)]\!]$.

\end{itemize}
   \\
    \hline
    \end{tabular}
    \caption{Detailed procedure of $\mathsf{SecDis}$.}
    \label{fig:SecDis}
\end{figure}

\textbf{\textit{SecSel:}} To identify gradients that can represent benign users, $S_1$ and $S_2$ jointly call decryption algorithm $\mathsf{PartDec1}$ and $\mathsf{PartDec2}$ to obtain $(sum_{i \in [1,n]})$ and find the $\lceil\frac{n}{2}\rceil$ gradients with the largest corresponding value in $(sum_i)$ among the $n$ gradients. To minimize the time spent in the selection process, we use the Randomized-Selection (RS) algorithm, which can solve the this top-k problem in $\mathcal{O}(n)$ time complexity.

% \begin{figure}[h]
%   \centering
  
%     \begin{tabular}{|p{\columnwidth}|}
%     \hline

% \begin{center} \textbf{Implementation of $\mathsf{SecSel}$}\end{center}

% \textbf{Input:} $S_1$ holds the ciphertext of sum of the distances between each gradient and the other gradients $([\![sum_i]\!])$ $(i\in [1,n])$ and the secret key $sk_1$; $S_2$ holds the secret key $sk_2$. 

% \textbf{Output:} $\lceil\frac{n}{2}\rceil$ representative gradients.

% \textbf{Procedure:}
% \begin{itemize}

% \item[$\bullet$] \textbf{@$S_1$:} For $i\in [1,n]$, $S_1$ partially decrypt the $[\![sum_i]\!]$ by calling $[sum_i]_{sk_1} \gets \mathsf{PartDec1}(sk_1, [\![sum_i]\!])$, and send {$[sum_i]_{sk_1}$} and {$[\![sum_i]\!]$} to $S_2$.

% \item[$\bullet$]\textbf{@$S_2$:} To fully decrypt $([\![sum_i]\!])$ $(i\in [1,n])$, $S_2$ calls algorithm $sum_i \gets \mathsf{PartDec2}(sk_2, [\![sum_i]\!], [sum_i]_{sk_1})$ . Then $S_2$ Calls the Randomized Select algorithm to find the largest $\lceil\frac{n}{2}\rceil$ Value in the array $(sum_i)$. Lastly, $S_2$ Select the $\lceil\frac{n}{2}\rceil$ gradients $[\![\mathbf{g}_i]\!]$ with the largest corresponding value in $(sum_i)$ as the representative gradients.
% \end{itemize}
%    \\
%     \hline
%     \end{tabular}
%     \caption{Detailed procedure of $\mathsf{SecSel}$.}
%     \label{fig:SecSel}

% \end{figure}

\textbf{\textit{SecRep:}} To supplement the defense vulnerability, PROFL removes outliers in each dimension using the Pauta criterion, also known as the $3\sigma$ criterion. The Pauta criterion is a very common statistical method for dealing with outliers. It can accurately rejects outliers and preserves benign value. Submitted gradients has a symmetrical distribution with high in the middle and low on both sides, which can be approximated as a normal distribution, thus it meets the preconditions of the Pauta criterion. Finally, for each dimension, servers re-statistics and select the median as the representative value of the current dimension, as the final result of the aggregation. To protect the privacy of gradients in the above process, we need to use blinding techniques to perform statistics on the decrypted gradient value without exposing the plaintext. The specified procedure is shown in Fig.\ref{fig:SecRep}.

\begin{figure}[h]
  \centering
  \small
    \begin{tabular}{|p{0.95\columnwidth}|}
    \hline

\begin{center} \textbf{Implementation of $\mathsf{SecRep}$}\end{center}

\textbf{Input:} $S_1$ holds the encrypted value of all gradients in the $t$-th dimension $d_t=\{[\![\mathbf{g}_1(t)]\!], [\![\mathbf{g}_2(t)]\!], ...[\![\mathbf{g}_{n'}(t)]\!] \}(n' = \lceil\frac{n}{2}\rceil)$ and the secret key $sk_1$; $S_2$ holds the secret key $sk_2$. 

\textbf{Output:} The representative of all gradients in the $t$-th dimension $med_t$.

\textbf{Procedure:}
\begin{itemize}
\item[$\bullet$] $@S_1$: To protect privacy, $S_1$ randomly selects a nonzero noise vector $\mathbf{r} \gets Z_N^*$ to blind the gradients $\mathbf{g}_{i}(t)$ $(i\in [1,n'])$ using Equation(\ref{he2}) such that $[\![\mathbf{g}_i(t)']\!]=[\![\mathbf{g}_i(t)]\!]\cdot[\![\mathbf{r}]\!]=[\![\mathbf{g}_i(t)+\mathbf{r}]\!]$. Then, $S_1$ partially decrypts the gradient by calling $[\mathbf{g}_i(t)']_{sk_1} \gets \mathsf{PartDec1}(sk_1, [\![\mathbf{g}_i(t)']\!])$, and send {$[\mathbf{g}_i(t)']_{sk_1}$} and {$[\![\mathbf{g}_i(t)']\!]$} to $S_2$.

\item[$\bullet$]$@S_2$: To fully decrypt the gradients $[\![\mathbf{g}_{i}(t)']\!]$ $(i\in [1,n'])$, $S_2$ calls $\mathbf{g}_i(t)' \gets \mathsf{PartDec2}(sk_2, [\![\mathbf{g}_i(t)']\!], [\mathbf{g}_i(t)']_{sk_1}).$ Then, $S_2$ calculates the mean $\mu$ and variance $\sigma$ of $\mathbf{g}_i(t)'$ and eliminates the gradient value outside the range of $\mu \pm 3\sigma$ according to the Pauta criterion. Finally, $S_2$ re-statistics the gradient value to find the median $med_t'$ and sends it to $S_1$.

\item[$\bullet$]$@S_1$: To obtain the median $med_t$, $S_1$ removes the blinding $r$ of the median $med_t'$ using $med_t=med_t'-r$.
\end{itemize}
   \\
    \hline
    \end{tabular}
    \caption{Detailed procedure of $\mathsf{SecRep}$.}
    \label{fig:SecRep}
\end{figure}

\subsection{Integrate and Distribute}

After executing the $\mathsf{SecRep}$ in each dimension, $S_1$ concatenates the representatives of each dimension to get the global gradient $\textbf{g}_{agg} = (med_1, ... med_m)$. Then, convert the global gradient from integer type to floating-point type and use it to update the current global model according to $\mathbf{w}':=\mathbf{w}-lr \cdot \mathbf{g}$. Finally, the latest global model $\mathbf{w}'$ is encrypted with the public key $pk_m$ using $[\![\mathbf{w}']\!] \gets \mathsf{Enc}(pk_m, \mathbf{w}')$, and distributed to each user.\\

\section{Theoretical Analysis}\label{Theoretical Analysis}

\subsection{Security Analysis}

We provide a standard hybrid argument so as to prove that the joint view of servers $S_1$ and $S_2$ does not disclose any sensitive information about local gradients during protocol execution. In other words, in the honest-but-curious setting, PROFL achieves IND-CPA security~\cite{paillier1999efficient}. The security theorem is shown in Theorem~\ref{thm:sec}, and due to space constraints, the proof is omitted here.

\begin{theorem}\label{thm:sec}
Given a security parameter $\kappa$, two non-colluding servers $S_1$ and $S_2$, users $U$, define a random variable $REAL^{U,\kappa}_{\Pi}$ to represent the joint view of servers $S_1$ and $S_2$ during the execution of the protocol $\Pi$. There exists a probabilistic polynomial-time (PPT) simulator $SIM$, whose view $SIM^{U,\kappa}_{\Pi}$ is computationally indistinguishable from $REAL^{U,\kappa}_{\Pi}$.
\end{theorem}

\subsection{Complexity Analysis}

To evaluate the efficiency of PROFL, we discuss the communication and computational complexity for each training iteration, as shown in Table \ref{tab:com}.

\begin{table}[h]\centering

\caption{Computation and communication complexity in PROFL}
\setlength{\tabcolsep}{1.5pt}
\begin{tabular}{|c|cc|}
\hline
\multirow{2}{*}{Phase} & \multicolumn{2}{c|}{PROFL}                   \\ \cline{2-3} 
                       & \multicolumn{1}{c|}{Compu.}         & Comm. \\ \hline
$\mathsf{SecDis^*}$                 & \multicolumn{1}{c|}{$\mathcal{O}(n^2m(T_{Enc}+T_{Add})+nm(T_{Dec}+T_{Mul}))$} &  $\mathcal{O}(2nm|X|)$     \\ \hline
$\mathsf{SecSel}$                 & \multicolumn{1}{c|}{$\mathcal{O}(nT_{Dec})$}               &   $\mathcal{O}(n)$    \\ \hline
$\mathsf{SecRep}$                 & \multicolumn{1}{c|}{$\mathcal{O}({n}{ m(T_{Add}+T_{Dec})})$}               &  $\mathcal{O}(2m|X|)$     \\ \hline

\multicolumn{3}{l}{Note: $\mathsf{SecDis^*}$ indicates that $\mathsf{SecDis}$ is executed multiple times.}
\end{tabular}
\label{tab:com}
\end{table}

Let $n$ and $m$ be the number of users participating in the aggregation and the number of dimensions of gradients. $|X|$ represents the communication complexity of an encrypted number. $T_{Add}$, $T_{Mul}$, $T_{Enc}$, and $T_{Dec}$ are the time complexities of homomorphic operations, where homomorphic decryption consists of partial decryption step one and two. The time complexities  $T_{Enc} > T_{Dec} >> T_{Mul} >  T_{Add}$.

When calculating the centrality of $n$ gradients, each gradient must calculate the Euclidean distance with the rest of the gradients, so $\mathsf{SecDis}$ needs to be called $\sum_{i=1}^{n-1} i$ times. However, in this process, many intermediate variables are repeatedly calculated, which can be avoided by storing them. The same approach can be taken for the communication. Therefore, the computational and communication complexity can be greatly reduced. It's worth noting that we observe a positive correlation between the computational complexity of the $\mathsf{SecDis}$ phase and $n^2$. However, considering that in practical applications, such as in the medical and banking fields, each user represents an institution, the quantity of users $n$ is often small. Therefore, the computational overhead of $\mathsf{SecDis}$ won't become unacceptable due to the growth of $n$.

In the $\mathsf{SecSel}$ protocol, the complexity mainly depends on the number of gradients $n$. In the $\mathsf{SepRep}$ process, the complexity is limited to the number of benign gradients $\lceil\frac{n}{2}\rceil$ and the gradient dimension $m$.

\section{Performance Evaluation}\label{Performance Evaluation}

In this section, we conducted experiments on two datasets using multiple different settings to evaluate the performance of PROFL: (1) evaluating the performance of PROFL under various attack settings (attack method, attack ratio); (2) comparing the performance of PROFL with four other typical defense schemes, including Krum~\cite{DBLP:conf/nips/BlanchardMGS17}, Trimmed Mean~\cite{yin2018byzantine}, PEFL~\cite{DBLP:journals/tifs/LiuLXCHL21}, and ShieldFL~\cite{ShieldFL}. To highlight the advantages of PROFL, we used FedAvg~\cite{mcmahan2017communication} without any defense as a baseline.

\subsection{Testbed and Methodology}

All experiments were run on a Macbook pro computer equipped with a 12-core and 16GB unified memory M2 pro processor based on ARM architecture. We implemented the two-trapdoor AHE protocol at the 80-bit security level ~\cite{catalano2001bit} using python 3.8.11. The PROFL algorithm was tested using the pytorch library and lightweight threads were used to simulate the local learning processes of multiple users.

\textbf{\textit{Data Setting:}} To evaluate the performance of PROFL, we conducted experiments on two benchmark datasets, MINIST and FashionMINIST. To simulate real situations, the local learning model of each user should be underfitting, which can reflect the value of FL. Therefore, we believe that the total number of training data should be set at about 10\%, which is 6,000. PROFL is designed for independently and identically distributed data, so we choose to randomly shuffle the dataset and distribute it equally to each user.

\textbf{\textit{Models and Hyperparameters:}} In this task, we use a classical logistic regression as the model structure, with the number of input and output are 784 and 10 respectively. We set the total number of users in the experiment to 20. For each experiment, we set the batch size to 256, momentum to 0.5, and perform 1000 iterations on datasets. The data point for each experiment is the average of 10 experimental results.

\textbf{\textit{Poisoning Attack Settings:}} To evaluate the ability of PROFL to resist poisoning attacks, we designed two types of attacks: target attacks~\cite{bagdasaryan2020backdoor} and non-target attacks. In addition, after the gradient generation of attacks, we added the insidious gradient attack described in the threat model. We simulate the process of a Byzantine adversary evade the robust defense method and poison the model by carefully designing the value of certain dimensions of the malicious gradient. To evaluate the performance of PROFL, we conducted experiments under different proportions of malicious users. We used attack ratio $Att_{ratio} =|U^*|/|U|$ to describe the percentage of malicious users, where $|U|$ is the number of all users and $|U^*|$ is the number of malicious users.

\subsection{Accuracy Evaluation}

Table \ref{tab:table1} measures the performance of PROFL on target and non-target attacks. In targeted attacks, we marked the difference in the testing accuracy of the global model for the specific source class $Acc_{source}$ between PROFL and baseline. In non-targeted attacks, we marked the difference in the testing accuracy $Acc$ between PROFL and baseline. $AI=Acc-Acc^*$ is the improvement in accuracy of the global model, where $Acc^*$ is the testing accuracy of the baseline model. $AI_{source}=Acc_{source}-Acc_{source}^*$ is the improvement in accuracy of the global model for a specific source class.

\begin{figure*}[h]
    \centering
    \subfigure[\small{Target attack.}]{\label{fig:b}\includegraphics[width=0.24\textwidth]{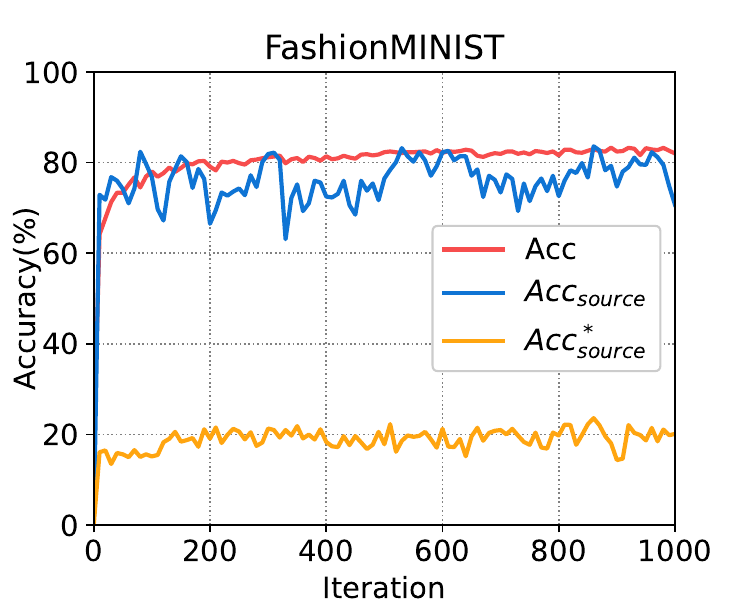}}
    \subfigure[\small Untarget attack under different $Att_{ratio}$.]{\label{fig:d}\includegraphics[width=0.24\textwidth]{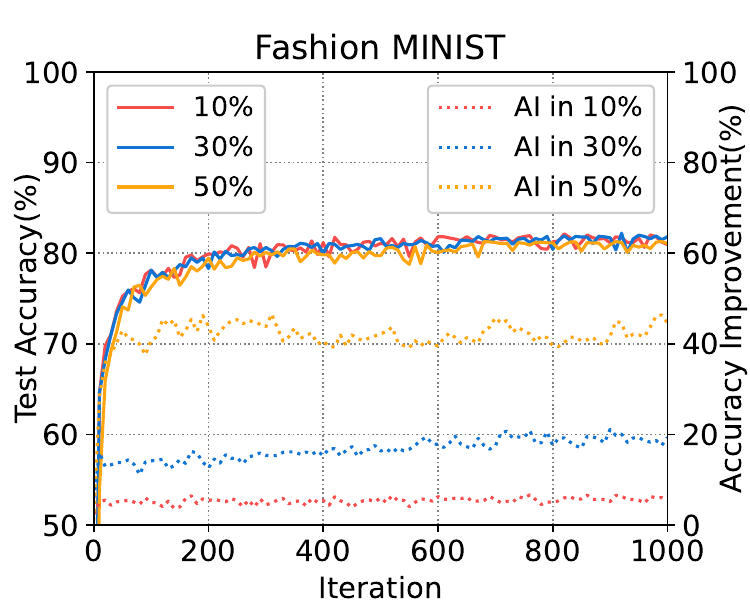}}
    \subfigure[\small Untarget attack under different defense strategies.]{\label{fig:e}\includegraphics[width=0.24\textwidth]{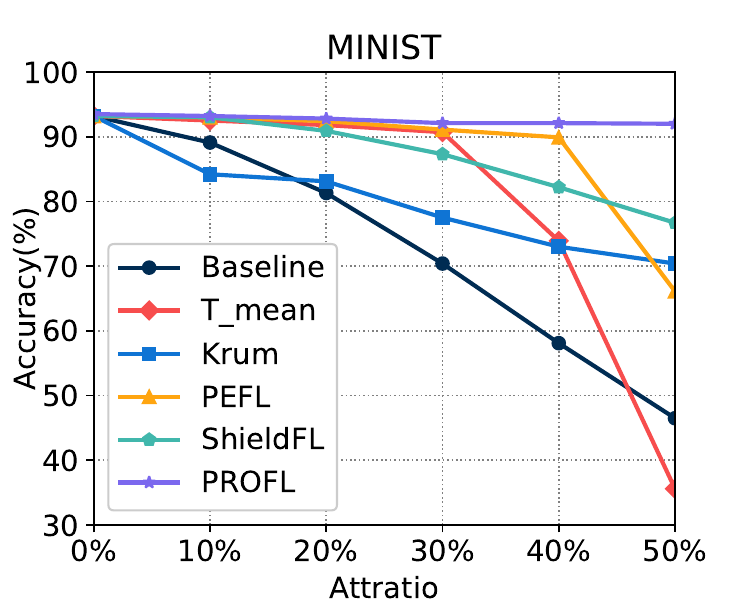}}
    \subfigure[\small Untarget attack under different defense strategies.]{\label{fig:f}\includegraphics[width=0.24\textwidth]{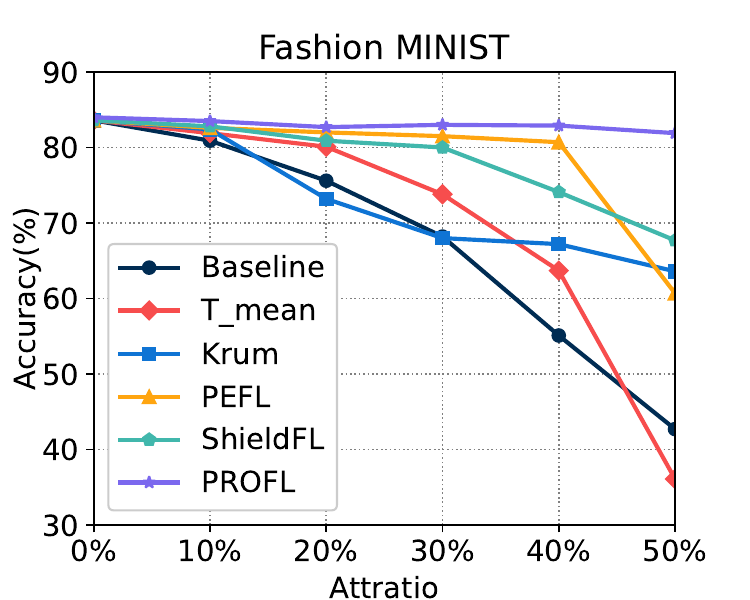}}
    \caption{EXPERIMENTAL RESULT.}
    \label{fig:abc}
\end{figure*}

\begin{table}[h]
\tiny
% \scriptsize
\centering
\caption{ACCURACY COMPARISON ($Att_{ratio}$ = 50\%)}
\label{tab:table1}
\begin{tabular}{|c|c|cc|cccc|}
\hline
\multirow{2}{*}{Datasets} &
  \multirow{2}{*}{Attack} &
  \multicolumn{2}{c|}{Baseline} &
  \multicolumn{4}{c|}{PROFL} \\ \cline{3-8} 
 &
   &
  \multicolumn{1}{c|}{$Acc^*_{s}$} &
  $Acc^*$ &
  \multicolumn{1}{c|}{$Acc_{s}$} &
  \multicolumn{1}{c|}{$Acc$} &
  \multicolumn{1}{c|}{$AI_{s}$} &
  $AI$ \\ \hline
\multirow{2}{*}{Minist} &
  Targeted &
  \multicolumn{1}{c|}{22.3\%} &
  $*$ &
  \multicolumn{1}{c|}{97.4\%} &
  \multicolumn{1}{c|}{92.8\%} &
  \multicolumn{1}{c|}{75.1\%} &
  $*$ \\ \cline{2-8} 
 &
  Untargeted &
  \multicolumn{1}{c|}{$*$} &
  46.5\% &
  \multicolumn{1}{c|}{$*$} &
  \multicolumn{1}{c|}{92.1\%} &
  \multicolumn{1}{c|}{$*$} &
  45.6\% \\ \hline
Fashion &
  Targeted &
  \multicolumn{1}{c|}{19.9\%} &
  $*$ &
  \multicolumn{1}{c|}{73.9\%} &
  \multicolumn{1}{c|}{82.3\%} &
  \multicolumn{1}{c|}{54.0\%} &
  $*$ \\ \cline{2-8} 
 Minist &
  Untargeted &
  \multicolumn{1}{c|}{$*$} &
  43.5\% &
  \multicolumn{1}{c|}{$*$} &
  \multicolumn{1}{c|}{82.5\%} &
  \multicolumn{1}{c|}{$*$} &
  39.0\% \\ 
  \hline
  \multicolumn{8}{l}{$Acc_{s}$ and $AI_{s}$ denotes $Acc_{source}$ ang $AI_{source}$, respectively.}
\end{tabular}
\end{table}
Fig.\ref{fig:b} show the accuracy trajectories with $Att_{ratio}$=50\% under the targeted attack, where the source class is `0' and the target class is `6'. As the figures indicate, PROFL has significantly improved $Acc_{sources}$ compared to the baseline.

To evaluate the impact of the ratio of attacks on PROFL, Fig.\ref{fig:d} depicts the change of $Acc$ and $AI$ with $Att_{ratio} \in [0\%, 30\%, 50\%]$. The image indicates that PROFL maintains stable accuracy and remarkable accuracy improvement at different $Att_{ratio}$. Even under the worst case scenario of $Att_{ratio}$ = 50\%, PROFL can still achieve high accuracy rates of up to 92\% and 82\% in the datasets MINIST and FashionMINIST  respectively, with accuracy improvements of 48\% and 43\%. This demonstrates the powerful robustness of PROFL.

As shown in Fig.\ref{fig:e}-\ref{fig:f}, We compared PROFL with existing excellent solutions at different $Att_{ratio}$. It can be observed that, as the proportion of malicious users increases, the accuracy of all defense measures except PROFL decreases significantly. We believe that this is because with the increase of $Att_{ratio}$, not only valuable data during the training process is reduced, but also the ability of defense methods to identify high-proportion malicious gradients weakens, leading to a deterioration of the robust aggregation effect. However, PROFL maintains excellent robustness throughout the process. We attribute this to PROFL's ability to accurately exclude poisoning gradients, making the contribution of benign gradients to the global gradient is purer.

\subsection{Communication and Computational Overhead}

We evaluated the communication overhead of PROFL. Additionally, we introduced the total communication volume provided by ShieldFL~\cite{ShieldFL}, which employs similar encryption approach as ours. This allows for a clear comparison with PROFL under the same security level. The communication volume comparison is illustrated in the Table.\ref{tab:comparison}. We can observe that PROFL demonstrates better control over communication overhead compared to ShieldFL. This is attributed to the fact that in PROFL, only during the $\mathsf{SecDis}$ process do servers transmit ciphertexts of gradients, while in other phases, data transmitted between servers consists of intermediate variables in plaintext or a small amount of encrypted data.

\begin{table}[]
\centering
\setlength{\tabcolsep}{2pt}
\caption{Total computation cost and running time of the defense strategy in one iteration}
\label{tab:comparison}
\begin{tabular}{|c|cccccc|}
\hline 
 & \multicolumn{6}{c|}{Total transmitted data (MB)} \\
                     \hline
Mothods                   
 & \multicolumn{1}{c|}{3 Users} & \multicolumn{1}{c|}{5 Users} & \multicolumn{1}{c|}{10 Users} & \multicolumn{1}{c|}{15 Users} & \multicolumn{1}{c|}{20 Users} & 25 Users \\  \hline
PROFL                    & \multicolumn{1}{c|}{18}      & \multicolumn{1}{c|}{31}      & \multicolumn{1}{c|}{63}       & \multicolumn{1}{c|}{96}       & \multicolumn{1}{c|}{129}      & 162      \\ \hline
ShieldFL                 & \multicolumn{1}{c|}{87}      & \multicolumn{1}{c|}{146}     & \multicolumn{1}{c|}{291}      & \multicolumn{1}{c|}{436}      & \multicolumn{1}{c|}{582}      & 728      \\ 
\hline
\hline
 & \multicolumn{6}{c|}{Running time}  \\
\hline
Mothods                   
 & \multicolumn{1}{c|}{3 Users} & \multicolumn{1}{c|}{5 Users} & \multicolumn{1}{c|}{10 Users} & \multicolumn{1}{c|}{15 Users} & \multicolumn{1}{c|}{20 Users} & 25 Users \\  \hline
PROFL                   & \multicolumn{1}{c|}{4.79 s}  & \multicolumn{1}{c|}{9.68 s}  & \multicolumn{1}{c|}{34.08 s}  & \multicolumn{1}{c|}{75.95 s}  & \multicolumn{1}{c|}{135.23 s} & 211.95 s \\
\hline
\end{tabular}
\end{table}

During the security defense process, we optimize task execution using 12-core multi-threading technique. We tested the average runtime of each algorithmic components in PROFL under different numbers of participants, as shown in the Table \ref{tab:comparison}. In the case of 20 participants, after optimizing scheduling, the runtime for one iteration of the privacy-preserving defense strategy is 34.08 seconds. In the same setup, ShieldFL provides data of around 1.2 seconds. However, it should be noted that this data is obtained from running on devices different from ours. Compared to ShieldFL, although PROFL does not exhibit significant advantages in terms of communication and computation overheads, it provide greater robustness and security to federated learning in scenarios with a higher proportion of poisoning attackers or instances of covert poisoning attacks.

% \begin{table}[h]
% \setlength{\tabcolsep}{2pt}
% \caption{Running time of the defense strategy in one iteration}
% \label{tab:tableCompute}
% \begin{tabular}{|c|cccccc|}
% \hline
% \multirow{2}{*}{Method} & \multicolumn{6}{c|}{Running time}                                                                                                                                      \\ \cline{2-7} 
%                         & \multicolumn{1}{c|}{3 Users} & \multicolumn{1}{c|}{5 Users} & \multicolumn{1}{c|}{10 Users} & \multicolumn{1}{c|}{15 Users} & \multicolumn{1}{c|}{20 Users} & 25 Users \\ \hline
% PROFL                   & \multicolumn{1}{c|}{4.79 s}  & \multicolumn{1}{c|}{9.68 s}  & \multicolumn{1}{c|}{34.08 s}  & \multicolumn{1}{c|}{75.95 s}  & \multicolumn{1}{c|}{135.23 s} & 211.95 s \\ \hline
% \end{tabular}
% \end{table}

\section{Conclusion}\label{Conclusion}

In this paper, we propose PROFL to defend against various poisoning attacks while protecting privacy. We use a composite algorithm based on the Multi-Krum algorithm and Pauta criterion to identify and eliminate the interference from Byzantine poisoning attacks. At the same time, we use the two-trapdoor AHE algorithm to protect the privacy of the entire FL process. Security analysis and experimental evaluation show that PROFL is superior to similar works in terms of security and robustness, while keeping communication and computation costs within an acceptable range. In future work, we will further explore methods to optimize efficiency and reduce the computational and communication costs of PROFL.

\end{document}